\documentclass[conference,9pt]{IEEEtran}

\usepackage{cite,spconf,amsmath,graphicx,hyperref}
\usepackage{bm}
\usepackage{amssymb}
\usepackage{algorithm}
\usepackage{algpseudocode}
\usepackage{xcolor}
\usepackage{subcaption}

\IEEEoverridecommandlockouts
\title{Single-Snapshot Localization Using Sparse Extremely Large Aperture Arrays}
\name{Yunqiao Hu$^{\dagger}$, Xuesu Xiao$^{\S}$, Steven Jones$^{\dagger}$, and Shunqiao Sun$^{\dagger}$
\thanks{ The work of Y.\ Hu and S.\ Sun was supported in part by U.S. National Science Foundation (NSF) under Grant ECCS-2340029 and Alabama Transportation Institute (ATI). }}
\address{$^{\dagger}$University of Alabama, Tuscaloosa, AL 35487 USA \\
$^{\S}$George Mason University, Fairfax, VA 22030 USA}
\begin{document}
%
\maketitle
\begin{abstract}
This paper investigates single-snapshot direction-of-arrival (DOA) estimation and target localization with coherent sparse extremely large aperture arrays (ELAAs) in automotive radar applications. Far-field and near-field signal models are formulated for distributed bistatic configurations. To enable noncoherent processing, a single-snapshot MUSIC (SS-MUSIC) algorithm is proposed to fuse local spectra from individual subarrays and extended to near-field localization via geometric intersection. For coherent processing, a single-snapshot ESPRIT (SS-ESPRIT) method with ambiguity dealiasing is developed to fully exploit the aperture of sparse ELAAs for high-resolution angle estimation. Simulation results demonstrate that SS-ESPRIT provides superior angular resolution for closely spaced far-field targets, while SS-MUSIC offers robustness in near-field localization and flexibility in hybrid scenarios.
\end{abstract}

\section{Introduction}
\label{sec:intro}
Sparse extremely large aperture arrays (ELAAs) have gained increasing attention in both communications and radar sensing due to their ability to form very large virtual apertures using spatially distributed and sparsely placed antenna elements \cite{10934790,9617121}. Compared with conventional uniform linear arrays (ULAs), they provide higher spatial resolution, greater angular diversity, and a wider field of view (FoV) with reduced hardware cost. In radar systems, the bi-/multi-static configurations of sparse ELAAs support coherent signal processing across widely separated nodes, improving detection and localization performance in dynamic environments.

The concept of sparse ELAA has been explored in next-generation wireless communications, such as 6G, under paradigms like extremely large-scale multiple-input multiple-output (XL-MIMO) array\cite{9617121,8644126,10496996}, enabling macrodiversity, high spectral efficiency, and robustness to blockage. In radar applications, the closest counterpart to ELAAs is the widely studied distributed aperture radar system, which has been proposed for early warning \cite{1706161}, surveillance \cite{8835491}, and radar networked sensing \cite{9107234,10879391}. These systems typically employ centralized control or coherent processing across multiple spatially separated modules to enhance the signal-to-noise ratio (SNR) and improve target localization accuracy.

Driven by the demands of autonomous driving, distributed radar concepts have recently been extended to the automotive domain \cite{EP3712653A1,9318740}. Deploying radar modules at various locations on a vehicle, such as bumper corners, the roof, or side mirrors, yields a sparse ELAA capable of supporting both far-field and near-field direction-of-arrival (DOA) estimation. This spatial diversity enables multi-view sensing, enhancing the detection performance of targets with angle-dependent radar cross sections (RCS)\cite{5393291,10422820,Spatial_diversity_EuRAD_2023}. Industrial solutions, such as Zendar's distributed aperture radar\cite{zedar_dar}, have demonstrated enhanced resolution and robustness at reduced power and cost compared to conventional four-chip imaging radars.
On the academic front, recent efforts have explored unified GLRT-based detection frameworks for distributed apertures \cite{7839239}, multimodal sensing with heterogeneous distributed MIMO radars leveraging Doppler beam sharpening beamforming for extended FoV and improved azimuth resolution \cite{10289419}.
\

Despite these advantages, DOA estimation and target localization using sparse ELAAs remain challenging in automotive scenarios. Many existing methods~\cite{s21227618,10548034} require multiple snapshots to estimate the signal subspace, making them unsuitable for highly dynamic environments where only a single snapshot is available~\cite{SUN_SPM_2020,Yunqiao_ICASSP_2024,Overview_ICASSP_2025}. Additionally, sparsity and large inter-module spacing introduce prominent sidelobes that can severely degrade estimation performance if not properly mitigated. Moreover, due to the large aperture size of ELAAs, near-field effects become significant and must be considered in the algorithm design. Recent works~\cite{10934790} have systematically reviewed the sensing models and classical methods for near-field localization with large-aperture arrays, and~\cite{10522654,10220205} provide comprehensive surveys on near-field DOA estimation and communication. However, limited attention has been given to near-field sensing challenges specific to sparse ELAA systems.
\

To the best of our knowledge, this work is the first to address the problem of single-snapshot DOA estimation and target localization with sparse ELAAs. We develop tailored far-field and near-field signal models for distributed bistatic radar configurations. To overcome the inherent limitations of single-snapshot data, we propose a Single-Snapshot MUSIC (SS-MUSIC) method that noncoherently fuses local spectra from separated subarrays. To further improve resolution and resolve ambiguities, we introduce a Single-Snapshot ESPRIT (SS-ESPRIT) algorithm equipped with an ambiguity-dealiasing strategy. Simulation results demonstrate that SS-ESPRIT achieves high-resolution DOA estimation for closely spaced far-field targets, while SS-MUSIC remains effective for near-field localization, thereby establishing the feasibility and robustness of single-snapshot processing in sparse ELAA-based radar systems.

\vspace{-1mm}

\section{System Model}
\label{system model}
\begin{figure}[t]
\begin{minipage}[b]{1.0\linewidth}
\centering
  \centerline{\includegraphics[width=8.0cm]{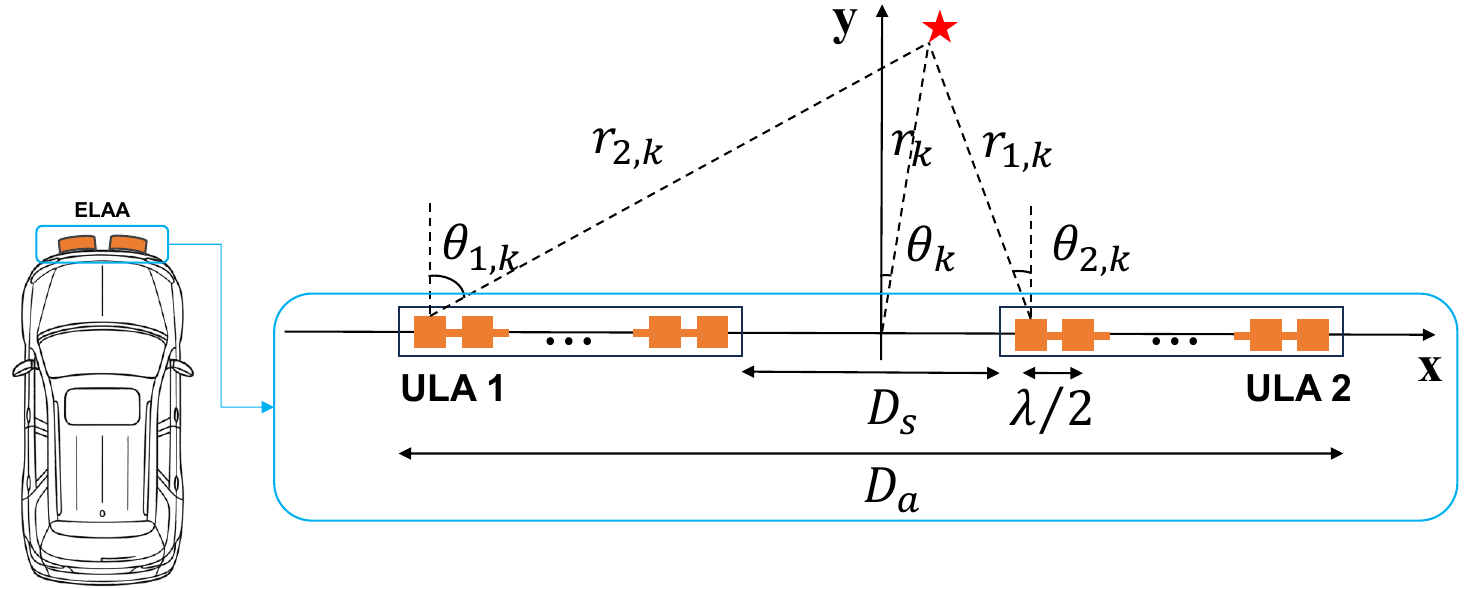}}
\end{minipage}
\caption{Illustration of a sparse ELAA composed of two uniform linear arrays (ULAs) symmetrically placed along the $x$-axis. }
\vspace{-4mm}
\label{dist_array_sketch}
\end{figure}

As illustrated in Fig.~\ref{dist_array_sketch}, consider a symmetric ELAA along the $x$-axis consist of two ULAs separated by distance $D_s \gg (M - 1)d$. Each ULA consists of $M$ elements with half-wavelength spacing $d = \frac{\lambda}{2}$, resulting in a total aperture of ELAA:
\[
D_a = D_s + 2(M - 1)d.
\]
Let $m \in \mathcal{M} = \{0, \ldots, M-1\}$ index the elements within each ULA, and $n \in \mathcal{N} = \{1, 2\}$ index the ULAs. The position of the $m$-th element in the $n$-th ULA is given by $\mathbf{u}_{n,m} = [x_{n,m},\, 0]^T$, where
\begin{equation}
x_{n,m} = \left(m - \tfrac{M-1}{2} \right)d + \left(n - \tfrac{3}{2} \right)\left(D_s + (M-1)d \right).
\end{equation}

Assume $K$ target sources located at
\begin{equation}
\mathbf{p}_k = r_k
\begin{bmatrix}
\cos\theta_k \\
\sin\theta_k
\end{bmatrix}, \quad k = 1,\dots,K,
\end{equation}
where $r_k$ and $\theta_k$ denote the range and DOA of the $k$-th target relative to the ELAA center.
The \((n,m)\)-th element of the steering vector $\mathbf{a}(r_k, \theta_k)$ for the $k$-th target is then expressed as
\begin{align}
a_{n,m}(r_k, \theta_k) &= \exp\left(-j \tfrac{2\pi}{\lambda} \left\|\mathbf{p}_k - \mathbf{u}_{n,m} \right\| \right) \nonumber \\
&= \exp\left(-j \tfrac{2\pi}{\lambda} \sqrt{r_k^2 - 2r_k x_{n,m} \cos\theta_k + x_{n,m}^2} \right).
\label{distributed_array_steer_vector}
\end{align}
To accurately perform DOA estimation with ELAA, it is critical to distinguish between far-field and near-field regimes and adopt the appropriate signal models for each.
\subsection{Far-Field Steering Vector for Sparse ELAAs}
\label{far_field_model}
When the target range satisfies $r_k \ge \frac{2D_a^2}{\lambda}$, the source lies in the far-field region of the entire distributed array. In this regime, the electromagnetic wavefront can be approximated as planar, and the propagation distance to the $(n,m)$-th sensor can be linearly approximated as:
\begin{align}
r_{n,m}^k \approx r_k - x_{n,m} \cos{\theta_k}.
\end{align}
The \((n,m)\)-th entry of the steering vector in (\ref{distributed_array_steer_vector}) is approximated as:
\begin{align}
a_{n,m}(r_k,\theta_k) \approx e^{-j\frac{2\pi}{\lambda} r_k} \cdot e^{j\frac{2\pi}{\lambda} x_{n,m} \cos{\theta_k}}, \quad n \in \mathcal{N},\; m \in \mathcal{M},
\end{align}
where the common phase term $e^{-j\frac{2\pi}{\lambda} r_k}$ can be omitted in DOA estimation.

\subsection{Near-Field Steering Vector for Sparse ELAAs}
\label{near_field_model}
For ranges $ \frac{2S_{a}^{2}}{\lambda} \le r_k< \frac{2D_{a}^{2}}{\lambda}$ the wavefront is spherical across the global ELAA but can be approximated as planar within each local ULA. Let $r_{n,0}^k$ denote the distance from the $k$-th target to the reference element ($m = 0$) of the $n$-th ULA:
\begin{align}
    r_{n,0}^{k} = \sqrt{r_{k}^{2}-2r_{k}x_{n,0}\cos{\theta_{k}}+x_{n,0}^{2}}, \forall n \in \mathcal{N}
\end{align}
where $x_{n,0} = (n-2)(M-1)d + (n - \tfrac{3}{2})D_s$.
with $x_{n,0}=(n-2)(M-1)d + (n-\frac{3}{2})D_s$. The $k$-th target DOA relative to the $n$-th ULA is:
\begin{align}
    \cos{\theta_{n,k}} = \frac{r_{k}\cos{\theta_{k}} - x_{n,0}}{r_{n,0}^{k}}
\end{align}
The resulting \((n,m)\)-th element of $\mathbf{a}(r_k,\theta_k)$ is:
\begin{align}
    a_{n,m}(r_k,\theta_k) \approx e^{-j\frac{2\pi}{\lambda} r_{n,0}^{k}} \cdot e^{j\frac{2\pi}{\lambda} m d \sin{\theta_{n,k}}}, \quad n \in \mathcal{N},\; m \in \mathcal{M},
    \label{distributed_array_steer_vector_nf}
\end{align}
For ranges satisfying
\begin{equation}
\max\left(5D_a, \tfrac{4D_aD_s}{\lambda}\right) \le r_k < \tfrac{2D_a^2}{\lambda},
\end{equation}
the DOA variations across ULAs become negligible\cite{li2023near,10545312}, i.e., $\theta_{n,k} \approx \theta_k$. Thus (\ref{distributed_array_steer_vector_nf}) further simplifies to
\begin{align}
    a_{n,m}(r_k, \theta_k) = e^{-j \frac{2\pi}{\lambda} r_{n,0}^{k}} \cdot e^{j \frac{2\pi}{\lambda} m d \sin{\theta_k}}, \quad n \in \mathcal{N},\; m \in \mathcal{M},
    \label{eq:steering_element_nf}
\end{align}
\subsection{Single Snapshot Signal Model for Sparse ELAAs}
Assuming $K$ independent stationary sources, the received signal for a single snapshot is modeled as
\begin{align}
\mathbf{x} = \sum_{k=1}^{K} s_k,\mathbf{a}(r_k,\theta_k) + \mathbf{n},
\end{align}
where $s_k$ denotes the RCS of the $k$-th source, $\mathbf{a}(r_k,\theta_k)$ is the steering vector corresponding to the source located at range $r_k$ and direction $\theta_k$, and $\mathbf{n}$ is additive white Gaussian noise.

\section{Localization  with Sparse ELAAs}
\label{sec:pagestyle}
This section provides the first exploration of target localization with sparse ELAAs across both far-field and near-field conditions.
\subsection{Far-Field DOA Estimation with Sparse ELAAs}
\begin{figure}[htb]
\begin{minipage}[b]{1.0\linewidth}
\centering
  \centerline{\includegraphics[width=7.0cm]{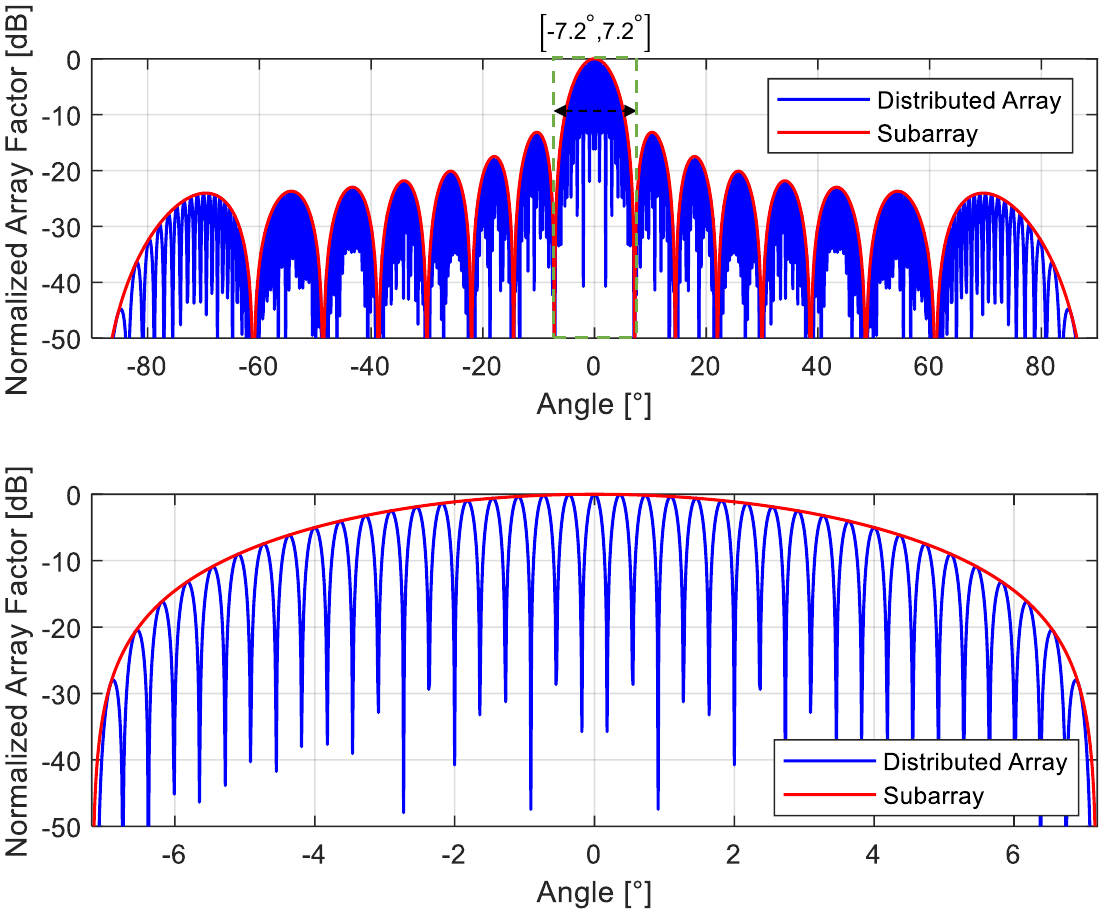}}
\end{minipage}
\caption{Array factor comparison between a sparse ELAA and its sub-ULA, with $D_s = 0.59~\mathrm{m}$ and $M = 16$. The lower plot shows a zoomed-in view over the angular region $[-7.2^{\circ}, 7.2^{\circ}]$.}
\label{Dis_Array_Fact}
\end{figure}

Distributed arrays with large inter-module spacing provide high angular resolution due to their extended aperture. However, this also leads to prominent sidelobes, as shown in Fig.~\ref{Dis_Array_Fact}, the envelope of the ELAA's array factor aligns with that of its sub-ULA.
We propose two single-snapshot far-field DOA estimation methods based on SS-MUSIC \cite{liao2016music} and SS-ESPRIT \cite{fannjiang2016compressive}. Given measurement vector $\mathbf{y}_n = [\mathbf{y}_{n,1}, \mathbf{y}_{n,2}]$, corresponding to two ULAs in the sparse ELAA, we construct Hankel matrices:
\begin{align}
   & \mathcal{H}(\mathbf{y}_{n,l}) = \begin{bmatrix}
 \mathbf{y}_{n,l}(0) & \mathbf{y}_{n,l}(1) & \cdots  & \mathbf{y}_{n,l}(M-L) \\
 \mathbf{y}_{n,l}(1) & \mathbf{y}_{n,l}(2) & \cdots  & \mathbf{y}_{n,l}(M-L+1)\\
 \vdots  &  \vdots & \ddots  & \vdots\\
 \mathbf{y}_{n,l}(L) & \mathbf{y}_{n,l}(L+1) & \cdots & \mathbf{y}_{n,l}(M)
\end{bmatrix}, \nonumber \\
& l \in \left \{ 1,2 \right \}, 1\le L < M
\end{align}
Assuming $K < M$, these Hankel matrices are low-rank. Applying Singular Value Decomposition (SVD) yields:
\begin{align}
     \mathcal{H}(\mathbf{y}_{n,l}) = \begin{bmatrix}
 \mathbf{U}_{1,l} & \mathbf{U}_{2,l}
\end{bmatrix} \mathrm{diag}\left(\sigma_{1,l},...,\sigma_{s,l},...\right)\begin{bmatrix}
\mathbf{V}_{1,l}^{H}
 \\
\mathbf{V}_{2,l}^{H}
\end{bmatrix}
\end{align}
where $\mathbf{U}_{1,l} \in \mathbb{C}^{(L+1)\times s},  \mathbf{U}_{2,l} \in \mathbb{C}^{(L+1)\times (L+1-s)}, \mathbf{V}_{1,l} \in \mathbb{C}^{(M-L+1)\times s}, \mathbf{V}_{2,l} \in \mathbb{C}^{(M-L+1)\times (M-L+1-s)}$, \( \mathbf{U}_{1,l} \) spans the signal subspace, and \( \mathbf{U}_{2,l} \) spans the noise subspace. SS-MUSIC constructs pseudospectra:
\begin{align}
S_{l}(\theta) = \dfrac{\lVert \mathbf{A}_l(\theta)\rVert_{2}}{\lVert \mathbf{U}_{2,l}^{H}\mathbf{A}_l(\theta)\rVert_{2}}, \quad l=1,2
\label{spectra_ss_music}
\end{align}
and outputs $\hat{S}(\theta)=\max \left\{S_l(\theta)\right\}$ (see Algorithm~\ref{alg1}).
\begin{algorithm}
\caption{SS-MUSIC for Far-Field DOA Estimation with ELAA}
\label{alg1}
\begin{algorithmic}[1]
\State \textbf{Input: } $\mathbf{y}_n = [\mathbf{y}_{n,1}, \mathbf{y}_{n,2}]$
\State Form $\mathcal{H}(\mathbf{y}_{n,1})$, $\mathcal{H}(\mathbf{y}_{n,2})$
\State Apply SVD to extract $\mathbf{U}_{2,1}$, $\mathbf{U}_{2,2}$
\State Compute $S_1(\theta)$ and $S_2(\theta)$ via Eq.~(\ref{spectra_ss_music})
\State \textbf{Output: } $\hat{S}(\theta)=S_1(\theta)S_2(\theta)$
\end{algorithmic}
\end{algorithm}
SS-MUSIC processes each ULA separately and combines DOA estimates noncoherently, failing to fully utilize the ELAA's large aperture. To address this, we propose SS-ESPRIT, which exploits the shift-invariant structure of the full ELAA. By selecting subarrays, it generates high-resolution yet ambiguous estimates, which are then resolved via a dealiasing strategy for accurate and unambiguous DOA estimation.
\begin{figure*}[htb]
\centering
    \begin{subfigure}[b]{0.24\textwidth}
        \includegraphics[width=\linewidth]{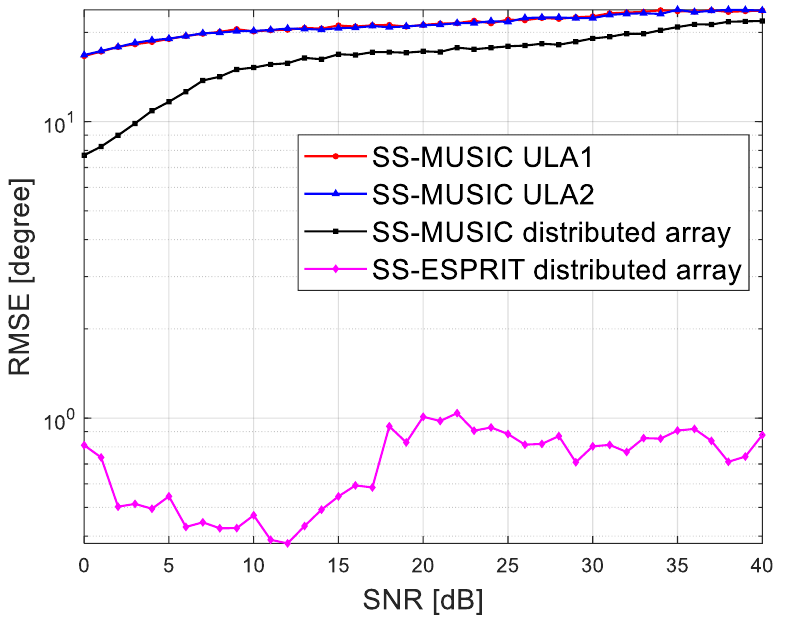}
        \caption{}
        \label{fig:rmse1}
    \end{subfigure}
    \hfill
    \begin{subfigure}[b]{0.24\textwidth}
        \includegraphics[width=\linewidth]{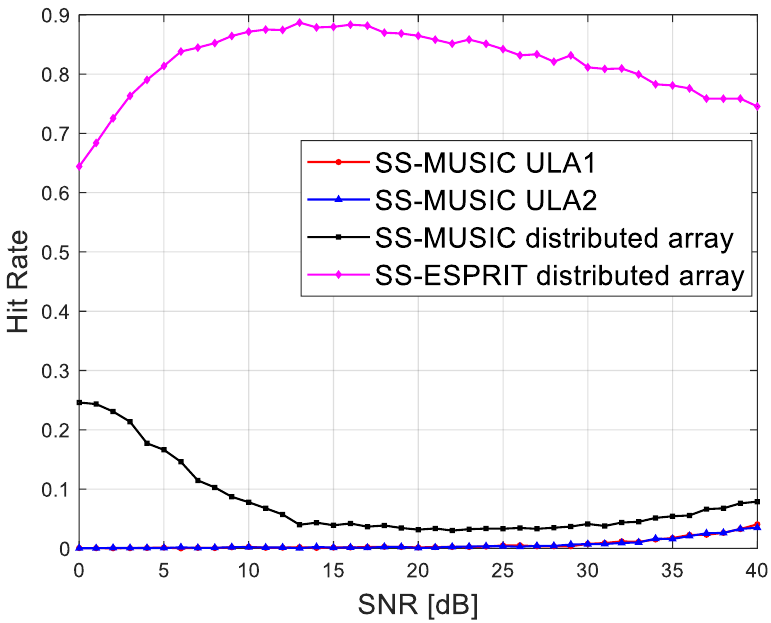}
        \caption{}
        \label{fig:hitrate1}
    \end{subfigure}
    \hfill
    \begin{subfigure}[b]{0.24\textwidth}
        \includegraphics[width=\linewidth]{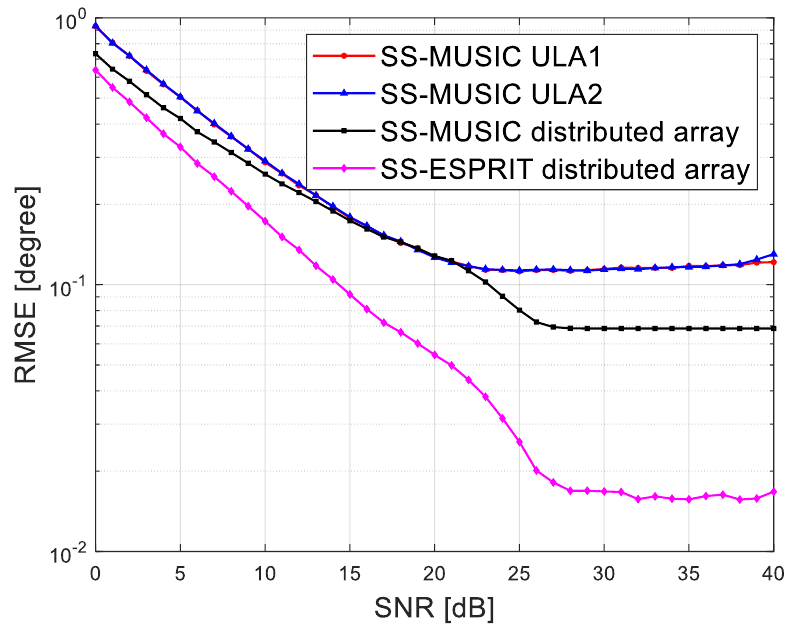}
        \caption{}
        \label{fig:rmse2}
    \end{subfigure}
    \hfill
    \begin{subfigure}[b]{0.24\textwidth}
        \includegraphics[width=\linewidth]{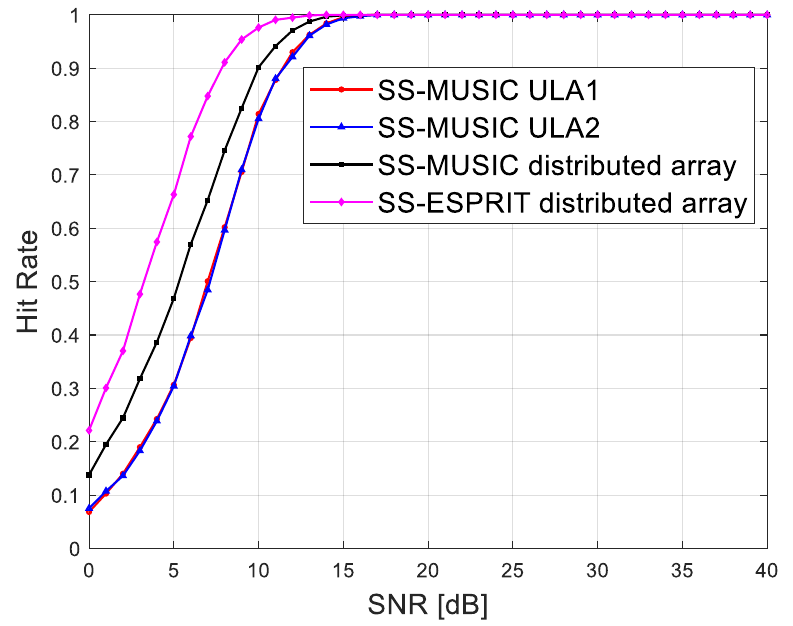}
        \caption{}
        \label{fig:hitrate2}
    \end{subfigure}
\caption{Performance of SS-MUSIC and SS-ESPRIT with ELAA: (a) RMSE, $0.4^{\circ}$ separation; (b) Hit rate, $0.4^{\circ}$ separation; (c) RMSE, $10^{\circ}$ separation; (d) Hit rate, $10^{\circ}$ separation.}
\vspace{-4mm}
\label{performance_evalue}
\end{figure*}
Two identical subarrays are extracted from the ELAA using selection matrices $\mathbf{J}_1$, $\mathbf{J}_2$:
\begin{align}
\mathbf{A}_{s1}(\bm{\theta}) = \mathbf{J}_1 \mathbf{A}_s(\bm{\theta}), \quad \mathbf{A}_{s2}(\bm{\theta}) = \mathbf{J}_2 \mathbf{A}_s(\bm{\theta})
\end{align}
with shift-invariance:
\begin{align}
\mathbf{J}_1 \mathbf{A}_s(\bm{\theta})\bm{\Phi} = \mathbf{J}_2 \mathbf{A}_s(\bm{\theta})
 \label{array_manifold_relation}
\end{align}
where $\mathbf{\Phi} = \mathrm{diag}(e^{j\phi_1}, e^{j\phi_2},...,e^{j\phi_K})$ is the phase shift matrix, and the phase shift $\phi_i = 2\pi \frac{\Delta_s}{\lambda}\mathrm{sin}(\theta_i)$, where $\Delta_s$ is the shift length between two subarrays. Following Step 1 to Step 3 in Algorithm~\ref{alg1}, the signal subspaces \( \mathbf{U}_{1,1} \) and \( \mathbf{U}_{1,2} \) corresponding to the two subarrays are obtained. Concatenating the signal subspaces from each ULA yields the full ELAA signal subspace \( \mathbf{U}_{1} = \begin{bmatrix} \mathbf{U}_{1,1} \\ \mathbf{U}_{1,2} \end{bmatrix} \in \mathbb{C}^{2M\times K}\). Assuming a nonsingular transformation \( \mathbf{T}_a \in \mathbb{C}^{K\times K}\), the signal subspace \( \mathbf{U}_1 \) and the array steering matrix \( \mathbf{A}_{s}(\bm{\theta}) \) satisfy:
\begin{align}
    \mathbf{U}_{1} = \mathbf{A}_{s}(\bm{\theta})\mathbf{T}_s
    \label{subspace_array_manifold_relation}
\end{align}
Substituting into the shift-invariance relation in (\ref{array_manifold_relation}) yields:
\begin{equation}
\mathbf{J}_1 \mathbf{U}_1 \bm{\Psi} = \mathbf{J}_2 \mathbf{U}_1,
\label{shift_invariant_array_manifold_relation}
\end{equation}
where \( \bm{\Psi} = \mathbf{T}_a^{-1} \bm{\Phi} \mathbf{T}_a \in \mathbb{C}^{K \times K} \). \( \bm{\Psi} \) and \( \bm{\Phi} \) are similar matrices and share the same eigenvalues. Define:
$\mathbf{U}_{1,1s} = \mathbf{J}_1 \mathbf{U}_1, \mathbf{U}_{1,2s} = \mathbf{J}_2 \mathbf{U}_1$.
Then, the equation in (\ref{shift_invariant_array_manifold_relation}) can be solved via the least squares (LS) method:
\begin{align}
    \hat{\bm{\Psi}} = (\mathbf{U}_{1,1s}^{H}\mathbf{U}_{1,1s})^{-1}\mathbf{U}_{1,1s}^{H}\mathbf{U}_{1,2s}
    \label{lsq_est}
\end{align}
The eigenvalues \( \{ \hat{\xi}_i \} \) of \( \hat{\bm{\Psi}} \) provide DOA estimates:
\begin{align}
    \hat{\theta}_i = \mathrm{arcsin} \left(\frac{\lambda\hat{\xi}_i}{2\pi \Delta_s} \right)
    \label{DOA_est_1}
\end{align}
When shift length $\Delta_s>\frac{\lambda}{2}$, the aliasing occurs\cite{765149,5307000}, and the candidate DOAs are $[\hat{\xi}_i ]_{i=1}^{K}$:
\begin{align}
    \hat{\theta}_{i}^{(q)} = \mathrm{arcsin}\left(\frac{\lambda\hat{\xi}_i}{2\pi \Delta_s}+q\frac{\lambda}{\Delta_s} \right)
    \label{DOA_est_2}
\end{align}
where the integer \( q \) satisfies:
\begin{align}
    \left \lceil \frac{\Delta_s}{\lambda}(-1-\hat{\xi}_i) \right \rceil \le q \le \left \lfloor \frac{\Delta_s}{\lambda}(1-\hat{\xi}_i) \right \rfloor, q \in \mathbb{Z} 
\end{align}
here $\lceil \cdot \rceil$ and $\lfloor \cdot \rfloor$ denote the ceiling and floor functions, respectively. By varying the shift length \( \Delta_s \) and applying consistency checks among candidate sets, the final unambiguous DOA estimates are obtained, as outlined in Algorithm~\ref{alg2}.

\begin{algorithm}
\caption{SS-ESPRIT for Far-Field DOA Estimation with ELAA}
\label{alg2}
\begin{algorithmic}[1]
\State \textbf{Input:} Measurement vector $\mathbf{y}_n = [\mathbf{y}_{n,1};\, \mathbf{y}_{n,2}] \in \mathbb{C}^{2M \times 1}$
\State Construct Hankel matrices $\mathcal{H}(\mathbf{y}_{n,1})$ and $\mathcal{H}(\mathbf{y}_{n,2}) \in \mathbb{C}^{(L+1)\times(M-L+1)}$
\State Perform SVD on each Hankel matrix to extract signal subspaces $\mathbf{U}_{1,1}$ and $\mathbf{U}_{1,2}$ (see Algorithm~\ref{alg1})
\State Form the concatenated signal subspace: $\mathbf{U}_1 = [\mathbf{U}_{1,1};\, \mathbf{U}_{1,2}] \in \mathbb{C}^{2M \times K}$
\For{each subarray shift $u$}
    \State Select subspaces $[\mathbf{U}_{1,1s}^{(u)}, \mathbf{U}_{1,2s}^{(u)}]$ using selection matrices $\mathbf{J}_1^{(u)}$, $\mathbf{J}_2^{(u)}$
    \State Estimate $\hat{\bm{\Psi}}^{(u)}$ via \eqref{lsq_est} and compute DOA candidates using \eqref{DOA_est_1} and \eqref{DOA_est_2}
\EndFor
\State Perform consistency check across candidate sets to resolve aliasing and obtain final DOA estimates
\end{algorithmic}
\end{algorithm}

\subsection{Near-Field Localization with Sparse ELAAs}
As discussed in Section~\ref{near_field_model}, targets within the near-field region of an ELAA exhibit varying ranges and DOAs across individual ULA modules. Specifically, when $\tfrac{2S_a^2}{\lambda} \le r_k < \tfrac{2D_a^2}{\lambda}$, the array steering vector follows the model in~\eqref{distributed_array_steer_vector_nf}. In this regime, each ULA can still adopt the far-field approximation locally, allowing  SS-MUSIC to be applied separately to extract DOA estimates.
Since the observed DOAs for the same target may vary across modules, cross-module association is required. We adopt the Matching Pursuit algorithm~\cite{258082} to pair corresponding estimates. Once associated, the target positions are computed by intersecting the two bearing lines defined by the associated DOAs and the known positions of the two ULA modules. For a given pair $(\hat{\theta}_{1,k}, \hat{\theta}_{2,k})$ corresponding to the $k$-th target, let $\mathbf{c}_1$ and $\mathbf{c}_2$ denote the first element positions of ULA 1 and ULA 2, respectively. Then, the bearing vectors from each array are:
\[
\mathbf{d}_1 = \begin{bmatrix} \cos(\hat{\theta}_{1,k}) \\ \sin(\hat{\theta}_{1,k}) \end{bmatrix}, \quad
\mathbf{d}_2 = \begin{bmatrix} \cos(\hat{\theta}_{2,k}) \\ \sin(\hat{\theta}_{2,k}) \end{bmatrix}
\]
The target position is estimated as the point of minimum distance between the two lines $\ell_1(r_{1,k}) = \mathbf{c}_1 + r_{1,k} \mathbf{d}_1$ and $\ell_2(r_{2,k}) = \mathbf{c}_2 + r_{2,k} \mathbf{d}_2$ in the least-squares sense. The overall procedure is summarized in Algorithm~\ref{alg3}.
\begin{algorithm}[htp]
\caption{SS-MUSIC for Near-Field DOA and Target Localization}
\label{alg3}
\begin{algorithmic}[1]
\State \textbf{Input: } Measurement vector $\mathbf{y}_n = [\mathbf{y}_{n,1};\, \mathbf{y}_{n,2}] \in \mathbb{C}^{2M \times 1}$
\State Construct Hankel matrices $\mathcal{H}(\mathbf{y}_{n,1})$ and $\mathcal{H}(\mathbf{y}_{n,2})$
\State Apply SVD to extract noise subspaces $\mathbf{U}_{2,1}$ and $\mathbf{U}_{2,2}$
\State Compute SS-MUSIC spectra and estimate local DOAs $\hat{\bm{\theta}}_1$, $\hat{\bm{\theta}}_2$
\State Associate $\hat{\bm{\theta}}_1$, $\hat{\bm{\theta}}_2$ using Matching Pursuit
\State Estimate target positions by intersecting bearing lines from each ULA
\State \textbf{Output: } Estimated target positions $[\hat{\mathbf{x}}_\text{target},\, \hat{\mathbf{y}}_\text{target}]$
\end{algorithmic}
\end{algorithm}

\section{Numerical Results}
\label{sec:typestyle}
\begin{figure}[htb]
\centering
    \begin{subfigure}[b]{0.48\linewidth}
        \includegraphics[width=\linewidth]{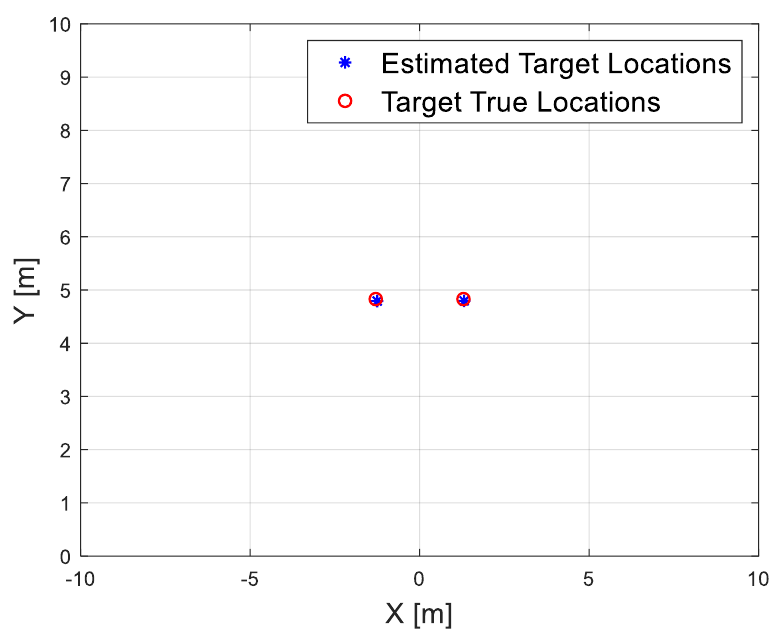}
        \caption{}
        \label{fig:nf_case1}
    \end{subfigure}
    \hfill
    \begin{subfigure}[b]{0.48\linewidth}
        \includegraphics[width=\linewidth]{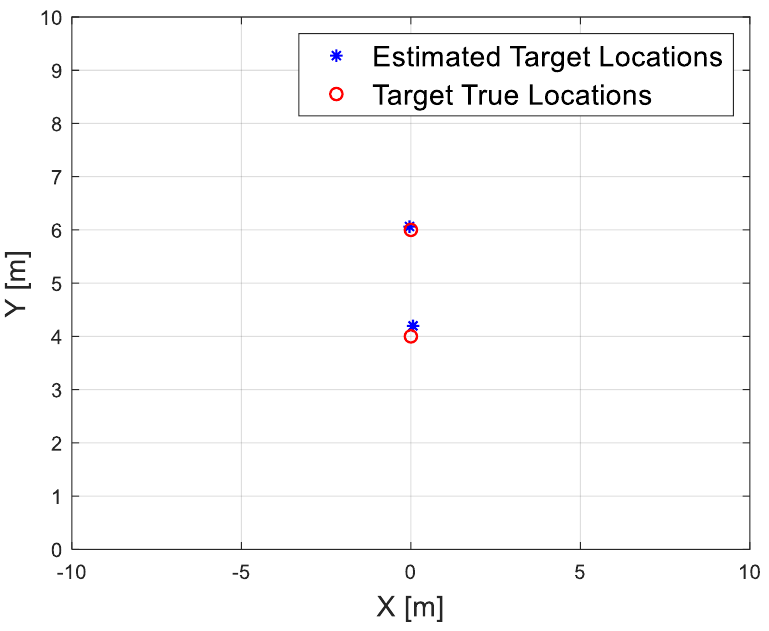}
        \caption{}
        \label{fig:nf_case2}
    \end{subfigure}
\caption{Estimated target positions in near-field scenarios.}
\label{fig:nf_sensing}
\end{figure}

We evaluate far-field DOA estimation performance using a sparse ELAA composed of two ULAs placed along the $x$-axis, each with $M = 16$ elements and inter-element spacing $d = \tfrac{\lambda}{2}$. Assuming a 76\,GHz operating frequency, the inter-module separation is $D_s = 150\lambda \approx 0.59\,\mathrm{m}$, yielding a Fraunhofer distance of approximately $\tfrac{2D_a^2}{\lambda} \approx 214.93\,\mathrm{m}$ (see Section~\ref{system model}). 
Two scenarios are considered: (1) \textit{Small angular separation}, where two targets at $r = 250\,\mathrm{m}$ (far-field) have DOAs $[-0.2^{\circ}, 0.2^{\circ}]$, with $0.4^{\circ}$ separation near the physical resolution limit ($0.3^{\circ}$) of the populated ELAA; (2) \textit{Large angular separation}, where the DOAs are $[-5^{\circ}, 5^{\circ}]$ ($10^{\circ}$ separation). In both cases, $5{,}000$ Monte Carlo trials are conducted over SNRs from $0$ to $40\,\mathrm{dB}$.
We evaluate performance using two metrics: the root mean squared error (RMSE), defined as:
\[
\mathrm{RMSE} = \sqrt{ \frac{1}{K N_\text{trial}} \sum_{n=1}^{N_\text{trial}} \sum_{k=1}^{K} \left( \hat{\theta}_k^{(n)} - \theta_k \right)^2 },
\]
and the hit rate, which is the percentage of trials where all DOA estimates fall within $\pm 0.5^\circ$ of the true values. We compare SS-MUSIC applied to the whole sparse ELAA and individual ULAs, as well as the proposed SS-ESPRIT method. As shown in Fig.~\ref{performance_evalue}(a)(b), when the target separation is small ($0.4^\circ$), SS-ESPRIT significantly outperforms SS-MUSIC, achieving higher hit rates and lower RMSE by leveraging the full ELAA aperture. For the large separation case ($10^\circ$) in Fig.~\ref{performance_evalue}(c)(d), both methods perform similarly, suggesting that single-ULA suffices for widely separated targets with SS-MUSIC. Overall, SS-MUSIC over the full ELAA shows slight improvement over per-ULA processing, while SS-ESPRIT offers notable advantages in resolving closely spaced sources. 
For near-field evaluation, Fig.~\ref{fig:nf_sensing}(a) two targets located at $r = 5\,\mathrm{m}$(i.e., within the near-field region $r < \max(5D_a, \tfrac{4D_aD_s}{\lambda}) \approx 19.5\,\mathrm{m}$), with DOAs of $[-10^{\circ}, 10^{\circ}]$. In Fig.~\ref{fig:nf_sensing}(b), two targets are located along the $y$-axis at $4\,\mathrm{m}$ and $6\,\mathrm{m}$, respectively. At a fixed SNR of $30\,\mathrm{dB}$, SS-MUSIC accurately resolves both cases, confirming its effectiveness for near-field localization under high-SNR conditions.

\section{Conclusions}
\label{sec:Conclusion}
This paper proposed single-snapshot DOA estimation and target localization algorithms tailored for distributed aperture radar systems operating in both far-field and near-field regimes. The SS-ESPRIT method exploits the extended aperture of the distributed array for coherent processing and achieves high-resolution angle estimation in the far-field. In contrast, the SS-MUSIC approach operates on individual subarrays and was further extended for near-field localization via geometric intersection of estimated DOAs. Simulation results demonstrate that SS-ESPRIT significantly outperforms SS-MUSIC in far-field scenarios with closely spaced sources, while SS-MUSIC remains effective in resolving near-field targets under high-SNR conditions. Notably, the SS-MUSIC framework offers flexibility in handling hybrid scenes containing both near-field and far-field sources, making it a promising candidate for practical automotive radar applications.

\newpage
\bibliographystyle{IEEEbib}
\bibliography{strings,refs}

\end{document}